\documentclass[aps,superscriptaddress,preprintnumbers,showpacs,onecolumn]{revtex4}
\usepackage{graphicx}
\usepackage{epsfig}
\usepackage{dcolumn}
\usepackage{amsmath}
\usepackage{bm}

\def\be{\begin{equation}}
\def\ee{\end{equation}}
\def\bea{\begin{eqnarray}}
\def\eea{\end{eqnarray}}

\begin{document}

\title{State-Relevant Maxwell's Equation from Kaluza-Klein Theory }

\author{Jing Luan}
\affiliation{Yuanpei College, Peking University, Beijing 100871,
China}
\author{Yongge Ma}\email{mayg@bnu.edu.cn}
\affiliation{Department of Physics, Beijing Normal University,
Beijing 100875, China}
\author{Bo-Qiang Ma}
\affiliation{School of Physics, Peking University, Beijing 100871,
China} \affiliation{MOE Key Laboratory of Heavy Ion Physics, Peking
University, Beijing 100871, China}

\begin{abstract}
We study a five-dimensional perfect fluid coupled with
Kaluza-Klein~(KK) gravity. By dimensional reduction, a modified form
of Maxwell's equation is obtained, which is relevant to the equation
of state of the source. Since the relativistic magnetohydrodynamics
(MHD) and the 3-dimensional formulation are widely used to study
space matter, we derive the modified Maxwell's equations and
relativistic MHD in 3+1 form. We then take an ideal Fermi gas as an
example to study the modified effect, which can be visible under
high density or high energy condition, while the traditional
Maxwell's equation can be regarded as a result in the low density
and low temperature limit. We also indicate the possibility to test
the state-relevant effect of KK theory in a telluric laboratory.

\end{abstract}


\maketitle

\renewcommand{\baselinestretch}{1.2}
\normalsize

\section{Introduction}
 A unified formulation of Einstein's theory
of gravitation and Maxwell's theory of electromagnetism in
four-dimensional (4D) spacetime was first proposed by Kaluza and
Klein using a five-dimensional (5D) geometry~\cite{Kaluza,Klein}. A
free test particle in 5D KK spacetime shows its electricity in the
reduced 4D spacetime when it moves along the fifth dimension.
Moreover a 5D dust field coupled with KK gravity can curve the 5D
spacetime in such a way that it provides exactly the source of the
electromagnetic field in the 4D spacetime after the
reduction~\cite{KK}. In this paper we study the coupling of a 5D
perfect fluid with KK gravity. It turns out that the 5D Einstein's
equation with a source gives a modification of Maxwell's equation
(see in section II), which can show its state-relevant effect on
high-density or high-temperature condition (see in section IV). Thus
this effect provides intriguing possibilities for the experimental
test of the KK theory. Note that the KK theory which we are
considering is purely classical. The KK theory has also been studied
from the particle physics point of view (see e.g.~\cite{Kubyshin}),
which can show its effects in high energy scale of $TeV$. Whereas
the energy scale needed for testing our state-relevant effect is
only around $keV$.

In order to reveal the physical implication of the modification more
clearly, both the modified Maxwell's equations and the corresponding
general relativistic MHD are reformulated in 3+1 form (see in
section III). The formalism is also useful for evolving numerically
a relativistic MHD fluid in a spacetime characterized by a strong
gravitational field. Taking an ideal Fermi gas as an example, the
modification term is studied as a function of degeneracy and
temperature parameters (see in section IV). Moreover the
modification terms for different components of the perfect fluid may
be different. This would result in a net charge excess in
high-temperature plasma. Recall that the the electrical neutrality
of atoms and of bulk matter has been examined precisely by a number
of experiments~\cite{Gillies}. However, in those experiments, either
the objects considered are not ionized, or the ions in the objects
cannot be regarded as a perfect fluid. Therefore, these experiments
cannot provide definite opposite evidence to the classical KK theory
since they do not satisfy our premise. But they do cast some doubts
on it. Taking account of the state-relevant character, we suggest
high-temperature plasma in earth laboratory or dense-matter white
dwarf in outer space as candidates to test the possible effects of
the modified Maxwell's equation.

\section{Kaluza-Klein gravity coupled with 5D perfect fluid}
Using a 5D geometry, Kaluza and Klein proposed a unified formulation
of gravity and electromagnetism in 4D spacetime~\cite{Kaluza,Klein}.
The original KK theory assumed the so-called``{\emph{Cylinder
Condition}}'', which means that there exists a space-like killing
vector field $\xi^a$ on the 5D spacetime ($\hat{M}, \hat{g}_{ab}$)
\cite{duff}-\cite{yang}. Note that the abstract index
notation~\cite{Wald} is employed throughout the paper and the
signature of the five-metric is of the convention $(-,+,+,+,+)$. In
addition, Kaluza also demanded that $\xi^a$ is normalized, {\it
i.e.},
\begin{equation}\label{phi}
\phi\equiv\hat{g}_{ab}\xi^a\xi^b=1\ .
\end{equation}
  Later research shows
that the Ansatz (\ref{phi}) may be dropped out and the $\phi$ may
play a key role in the study of
cosmology~\cite{uzan,BR,Wehus,Mohamm}. Being an extra dimension, the
orbits of $\xi^a$ are geometrically circles. The physical
consideration that any displacement in the usual ``physical'' 4D
spacetime (denoted as $M$) should be orthogonal to the extra
dimension implies that the ``{physical}'' 4D metric should be
defined as
\begin{equation}
g_{ab}=\hat{g}_{ab}-\phi^{-1}\xi_a\xi_b\ ,
\end{equation}
and the projection operator onto $M$ is
\begin{equation}
g^a_{~b}=\hat g^a_{~b}-\phi^{-1}\xi^a\xi_b\ .
\end{equation}
For practical calculation, it is convenient to take a coordinate
system $\{z^M=(x^\mu, y)|\mu=0,1,2,3\}$ with coordinate basis
$(e_M)^a=\{(e_\mu)^a,(e_5)^a\}$ on $\hat{M}$ adapted to $\xi^a$,
i.e., $(e_5)^a=(\frac{\partial}{\partial y})^a=\xi^a$. Then the
5-metric components $\hat{g}_{MN}$ take the form
\begin{equation}
\hat{g}_{MN}= \left(%
\begin{array}{cc}
  g_{\mu\nu}+\phi B_\mu B_\nu & \phi B_\mu \\
  \phi B_\nu & \phi \\
\end{array}%
\right),
\end{equation}
where $\hat{g}_{\mu5}\equiv\phi B_\mu$. So, locally, the ``physical"
spacetime can be understood as a 4-manifold $M$ with the coordinates
$\{x^\mu\}$ endowed with the metric $g_{ab}$. The whole theory is
governed by the 5D Einstein-Hilbert action
\begin{equation}\label{gaction}
S_G=-\frac 1{2\hat{k}}\int_{\hat{M}}d^4xdy\sqrt{-\hat{g}}\hat{R}\ .
\end{equation}
Suppose the range of the fifth coordinate to be $0\leq y\leq L$ and
denote $k=\hat{k}/L$. Let $B_\mu=fA_\mu, f^2=2k$, then equation
(\ref{gaction}) becomes a coupling action on $M$ as
\begin{equation}
\hat{S}_G=\int_M d^4x\sqrt{-g}\sqrt{\phi}\left(-\frac 1{2k}R+\frac 14\phi
F_{ab}(A)F^{ab}(A)\right)\ ,
\end{equation}
where $R$ is the curvature scalar of $g_{ab}$ on $M$ and
$F_{ab}(A)\equiv 2\partial_{[a}A_{b]}$. Thus, it results in a 4D
gravity $g_{ab}$ coupled to an electromagnetic field $A_a$ and a
scalar field $\phi$. It is clear that, under the Ansatz (\ref{phi}),
5D KK theory unifies the Einstein's gravity and the source-free
Maxwell's field in the standard formulism.

Now we consider a 5D perfect fluid
\begin{equation}
\hat T_{ab}=(\hat p+\hat \mu)\hat V_a\hat V_b+\hat p\hat g_{ab}\ .
\end{equation}
The 5-velocity $\hat V^a$ can be projected onto the ``{physical}''
 spacetime $(M, g_{ab})$ as
\begin{equation}\label{u}
u^a\equiv g^a_{~b}\hat V^b=\hat V^{\mu}(e_{\mu})^a-(B_{\mu}\hat
V^{\mu})(e_5)^a\ .
\end{equation}
Note that we have $\hat V^a\hat V_a=-1$, hence it is easy to show
that
\begin{equation}\label{umo}
\hat V^{\mu}\hat V_{\mu}\equiv \hat g_{ab}u^au^b=g_{ab}\hat
V^{a}\hat V^{b}=-1-\frac{Q^2}{\phi}\ ,
\end{equation}
 where $Q\equiv\hat
V_5=\hat g_{5a}\hat V^a$ represents the electric charge in $M$ \cite{KK}. The
energy-momentum tensor can be projected on M as $\tilde T_{ab}\equiv
g_a^{~c}g_b^{~d}\hat T_{cd}$. In order to obtain the observed 4D
energy-momentum tensor $T_{ab}$ on $M$, we have to integrate $\tilde
T_{ab}$ along the extra dimension. In the light of (\ref{u}) and
(\ref{umo}) we obtain
\begin{equation}
T_{ab}=(\mu+p)v_a v_b+pg_{ab}\ ,
\end{equation}
where
\begin{eqnarray}\label{relation}
p&=&\hat p\sqrt{\phi}L,\nonumber\\
\mu&=&\frac{\hat \mu L(Q^2+\phi)}{\sqrt{\phi}}+\hat p L\frac{Q^2}{\sqrt{\phi}},\nonumber\\
v_a&=&\frac{u_a}{\sqrt{-\hat V^{\mu}\hat V_{\mu}}}\ .
\end{eqnarray}
It is clear that $T_{ab}$ is the energy-momentum tensor of a 4D
perfect fluid in $M$, where $\mu$ and $p$ are respectively the
energy density and pressure density observed by a comoving observer
in $M$.

We now consider the reduction of 5D Einstein's equation
\begin{equation}
\hat R_{ab}-\frac{1}{2}\hat g_{ab}\hat R=\hat k\hat T_{ab}\ ,
\end{equation}
which is equivalent to
\begin{equation}\label{Ricci}
\hat R_{ab}=\hat k (\hat T_{ab}-\frac 13 \hat T^c_c \hat g_{ab})\ .
\end{equation}
It is not difficult to show from Eq.\ (\ref{gaction}) that the
components of the 5D Ricci tensor $\hat R_{ab}$ can be expressed as
\cite{Wehus}
\begin{equation}\label{doublefive}
\hat{R}_{55}=\frac 12 k\phi^2F^{\sigma\rho}F_{\sigma\rho}-\frac 12
\nabla^\mu\nabla_\mu\phi+\frac
1{4\phi}(\nabla^\mu\phi)\nabla_\mu\phi\ ,
\end{equation}
\begin{eqnarray}\label{five}
\hat{R}_{\mu 5}&=&\frac f2(\phi\nabla^\nu F_{\mu\nu}+\frac 32
F_{\mu\nu}\nabla^\nu\phi)\nonumber\\
&& + B_\mu\left(\frac 12 k\phi^2F^{\sigma\rho}F_{\sigma\rho}-\frac
12 \nabla^\nu\nabla_\nu\phi+\frac
1{4\phi}(\nabla^\nu\phi)\nabla_\nu\phi\right)\ ,
\end{eqnarray}
\begin{eqnarray}\label{nofive}
\hat{R}_{\mu\nu}&=&R_{\mu\nu}-k\phi F^\sigma_{\ \mu}
F_{\sigma\nu}-\frac 1{2\phi} \nabla_\mu\nabla_\nu\phi+\frac
1{4\phi^2}(\nabla_\mu\phi)\nabla_\nu\phi \nonumber \\
&&+B_\mu B_\nu\left(\frac 12
k\phi^2F^{\sigma\rho}F_{\sigma\rho}-\frac 12
\nabla^\sigma\nabla_\sigma\phi+\frac
1{4\phi}(\nabla^\sigma\phi)\nabla_\sigma\phi\right) \nonumber \\
&&+\frac f2 B_\mu(\phi\nabla^\sigma F_{\nu\sigma}+\frac 32
F_{\nu\sigma}\nabla^\sigma\phi)\nonumber \\
 &&+\frac f2 B_\nu(\phi\nabla^\sigma F_{\mu\sigma}+\frac 32
F_{\mu\sigma}\nabla^\sigma\phi)\ ,
\end{eqnarray}
where $\nabla _a$ is the 4D covariant derivative operator associated with $g_{ab}$.
Substituting Eq.\ (\ref{doublefive}) into Eq.\ (\ref{Ricci}), we
obtain a coupling equation for the matter fields as
\begin{equation}\label{df}
\frac{1}{2}k\phi^2F^{ab}F_{ab}=\sqrt{\phi}\nabla^a\nabla_a\sqrt{\phi}
+k\sqrt{\phi}\mu\left(1-\frac{2\phi}{3(\phi+Q^2)}\right)+k\sqrt{\phi}
p\frac{Q^2-\phi}{3(Q^2+\phi)}\ .
\end{equation}
Substituting Eq.\ (\ref{five}) into Eq.\ (\ref{Ricci}) and using
Eq.\ (\ref{df}), we obtain an electromagnetic field equation with
source as
\begin{equation}\label{f}
\phi\nabla^bF_{ab}+\frac32F_{ab}\nabla^b\phi=\tilde\gamma
(1+\frac{p}{\mu})J_a\ ,
\end{equation}
here we have defined $\tilde\gamma\equiv\sqrt{(1+Q^2)/(\phi+Q^2)}$,
$\rho\equiv\frac{ f\mu Q}{\sqrt{\phi(1+Q^2)}}$  and $J^a\equiv\rho
v^a$
 \cite{KK}. Substituting Eq.\ (\ref{nofive}) into Eq.\ (\ref{Ricci})
 and using Eq.\ (\ref{df}) and (\ref{f}),
we obtain a 4D Einstein's equation with source as
\begin{eqnarray}\label{nf}
G_{ab}&=&\frac{k}{\sqrt{\phi}}\left((\mu+p)v_a
v_b+g_{ab}p+\phi^{3/2}(F^{~c}_aF_{bc}-\frac14 g_{ab}F^{cd}F_{cd})
\nonumber \right. \\
 && \left. -\frac{1}{k}(g_{ab}\nabla^c \nabla_c
\sqrt\phi-\nabla_a\nabla_b \sqrt\phi)\right) ,
\end{eqnarray}
where $G_{ab}$ is the Einstein tensor of $g_{ab}$. More generally,
if the 5D perfect fluid consists of $m$ components, $\hat T_{ab}$
then reads
\begin{equation}
\hat T_{ab}=\sum_{\eta=1}^m\left((\hat p_\eta+\hat\mu_\eta)\hat
V_a(\eta)\hat V_b(\eta)+\hat p_\eta\hat g_{ab}\right)\ .
\end{equation}
By similar calculations, Eqs.
(\ref{df})-(\ref{nf}) become respectively
\begin{eqnarray}
\frac{1}{2}k\phi^2F^{ab}F_{ab}&=&\sqrt{\phi}\nabla^a\nabla_a\sqrt{\phi}
+k\sqrt{\phi}\sum_{\eta=1}^m\mu(\eta)\left(1-\frac{2\phi}{3(\phi+Q(\eta)^2)}\right)\nonumber\\
&&+k\sqrt{\phi}\sum_{\eta=1}^m
p(\eta)\frac{Q(\eta)^2-\phi}{3(Q(\eta)^2+\phi)}\ ,
\end{eqnarray}
\begin{equation}
\phi\nabla^bF_{ab}+\frac32F_{ab}\nabla^b\phi=\sum_{\eta=1}^m\tilde\gamma(\eta)
\left(1+\frac{p(\eta)}{\mu(\eta)}\right)J_a(\eta)\ ,
\end{equation}
\begin{eqnarray}
G_{ab}&=&\frac{k}{\sqrt{\phi}}\left(\sum_{\eta=1}^m((\mu(\eta)+p(\eta))v_a(\eta)
v_b(\eta)+g_{ab}p(\eta))+\phi^{3/2}(F^{~c}_aF_{bc}-\frac14
g_{ab}F^{cd}F_{cd}\right.\nonumber \\
&&\left.-\frac{1}{k}(g_{ab}\nabla^c \nabla_c
\sqrt\phi-\nabla_a\nabla_b \sqrt\phi)\right)\ .
\end{eqnarray}

It is interesting to see the results when $\phi\equiv 1$. Eqs.\
(\ref{df})-(\ref{nf}) become respectively
\begin{equation}
\frac12
kF^{ab}F_{ab}=k\mu\left(1-\frac{2}{3(1+Q^2)}\right)+kp\frac{Q^2-1}{3(Q^2+1)}\
, \end{equation}
\begin{equation}\label{Max}
\nabla^b F_{ab}=(1+\frac{p}{\mu})J_a \ ,
\end{equation}
\begin{equation}\label{Ein}
G_{ab}=k(T_{ab}^{({\mathrm{fluid}})}+T_{ab}^{({\mathrm{em}})}) \ ,
\end{equation}
 where
$T_{ab}^{({\mathrm{fluid}})}\equiv (\mu+p)v_av_b+pg_{ab}$ and
$T_{ab}^{({\mathrm{em}})}=F_a^{~c}F_{bc}-\frac14 g_{ab}F^{cd}F_{cd}$
are respectively the usual energy-momentum tensors of 4D perfect
fluid and electromagnetic field. Eq.~(\ref{Ein}) is the standard 4D
Einstein's equation while Eq.~(\ref{Max}) is not the same as the
standard 4D Maxwell's equation $\nabla^bF_{ab}=J_a$. The new term
$(1+\frac{p}{\mu})$ brings an effective charge which can be
considered as a state-relevant effect, as will be discussed later.
We thus call Eq.~(\ref{Max}) the state-relevant Maxwell's equation.

\section{Maxwell's equations and relativistic MHD in 3+1 form}
Since relativistic MHD is widely used to study space matter and 3D
formulation is frequently applied to dealing with specific
issues~\cite{Baumgarte}, we now derive the modified results of
Maxwell's equations and relativistic MHD in 3+1 form. The 4D
spacetime $M$ is foliated into a family of non-intersecting
spacelike three-surfaces $\Sigma$, which arise, at least locally, as
level surfaces of a scalar time function $t$. The spatial metric
$\gamma_{ab}$ on the three-dimensional hypersurfaces $\Sigma$ is
induced by the spacetime metric $g_{ab}$ according to
 \begin{equation}\label{first}
\gamma_{ab}=g_{ab}+n_an_b\ ,
 \end{equation}
 where $n^a$ is the unit normal vector to the slices and thus $n_a=-\alpha\nabla_at$. Here the
normalization factor $\alpha$ is called the lapse function. The time
vector $t^a$ is dual to the foliation 1-form $\nabla_at$ and can be
decomposed as
\begin{equation}
t^a=\alpha n^a+\beta^a \ , \end{equation}
 where the shift vector
$\beta$ is spatial, {\it i.e.}, $n_a\beta^a=0$. Since the extrinsic
curvature $K_{ab}$ of $\Sigma$ can be written as
\begin{equation}
K_{ab}=-\nabla_an_b-n_aa_b \ , \end{equation}
 where $a_a\equiv
n^b\nabla_b n_a$, the divergence of $n^a$ satisfies
\begin{equation}
\nabla_a n^a=-K \ ,\end{equation}
 here $K$ is the trace of $K_{ab}$.

Firstly we write the modified Maxwell's equations in 3+1 form. The
Faraday tensor $F^{ab}$ can be decomposed as
\begin{equation}
F^{ab}=n^a E^b-n^b E^a +\epsilon^{abc}B_c \ ,\end{equation} where
$E^a$ and $B^a$ are the electric and magnetic fields observed by a
normal observer $n^a$. Both fields are purely spatial, whereby
\begin{equation}
 E^an_a=0\ \ and\ \ B^an_a=0 \ ,
 \end{equation}
 and the three-dimensional Levi-Civita symbol $\epsilon_{abc}$
is defined by
\begin{equation}\label{seventh}
\epsilon_{abc}=n^d\epsilon_{dabc}\ \ or\ \
\epsilon^{abc}=n_d\epsilon^{dabc}\ .
\end{equation}
The electromagnetic current four-vector $J^a$ is decomposed as
\begin{equation}
J^a=n^a\rho_e +j^a \ ,\end{equation}
 where $\rho_e$ and $j^a$ are
the charge density and 3-current as observed by a normal observer
$n^a$. Note that $j^a$ is purely spatial, {\it i.e.}, $j^an_a=0$.
With these definitions, the modified Maxwell's equations (\ref{df}),
(\ref{f}) and
$\nabla_{[a}F_{bc]}=0$  can be
cast into 3+1 form as
\begin{eqnarray}
k\phi^2(B^2-E^2)&=&\sqrt{\phi}(D^aD_a\sqrt{\phi}-(\alpha^{-1}(\partial_t-\mathcal{L}_{\beta}))^2\sqrt{\phi}\nonumber\\
&&+K\alpha^{-1}(D_a\sqrt{\phi})(D^a\ln\alpha)(\partial_t-\mathcal{L}_{\beta})\sqrt{\phi}
)\nonumber\\
&&+k\sqrt{\phi}\mu\left(1-\frac{2\phi}{3(\phi+Q^2)}\right)+k\sqrt{\phi}p\frac{Q^2-\phi}{3(\phi+Q^2)},
\end{eqnarray}
\begin{eqnarray}
D_aE^a&=&\phi^{-1}(\tilde\gamma(1+\frac{p}{\mu})
\rho_e-\frac32E^aD_a \phi),\label{M1} \\
\phi\mathcal{L}_tE^a&=&\phi(\alpha
KE^a+\mathcal{L}_{\beta}E^a+\epsilon^{abc}D_b(\alpha B_c ))
\nonumber\\
&&-\alpha\tilde\gamma(1+\frac{p}{\mu})j^a+\frac32\alpha(\epsilon^{abc}B_bD_c\phi
-E^a\alpha^{-1}(\partial_t-\mathcal{L}_{\beta})\phi)\ , \label{M2}\\
D_aB^a&=&0 \ ,\\
\mathcal{L}_tB^a&=&-\epsilon^{abc}D_b(\alpha E_c)+\alpha
KE^a+\mathcal{L}_\beta B^a \ , \end{eqnarray} here
$\mathcal{L}_\beta$ denotes the Lie-derivative along $\beta^a$ and
$D_a$ is the covariant derivative operator associated to
$\gamma_{ab}$. Note that the Lie-derivative of a spacelike tensor
$A^{a\cdots b}_{~c\cdots d}$ along $s^a$ is defined conventionally
as $\tilde{\mathcal{L}}_sA^{a\cdots b}_{~c\cdots d}\equiv
\gamma^a_e\cdots \gamma^b_f\gamma^g_c\cdots
\gamma^h_d\mathcal{L}_sA^{e\cdots f}_{~g\cdots h}$, and we write
$\tilde{\mathcal{L}}_s$ as $\mathcal{L}_s$ for short. Note also that
the formula $n^b\nabla_bn_a=a_a=D_aln\alpha$ is used in the above
calculation~\cite{1982}. If one considered a 5D perfect fluid
consisting of $m$ components, the terms
$\tilde\gamma(1+\frac{p}{\mu}) \rho_e$ and
$\tilde\gamma(1+\frac{p}{\mu})j^a$ in Eqs.~(\ref{M1}) and (\ref{M2})
would be replaced by $\sum_{\eta=1}^m
\tilde\gamma(\eta)(1+\frac{p(\eta)}{\mu(\eta)}) \rho_e(\eta)$ and
$\sum_{\eta=1}^m
\tilde\gamma(\eta)(1+\frac{p(\eta)}{\mu(\eta)})j^a$. When
$\phi\equiv 1$, one can see from Eq.~(\ref{M1}) that $\tilde
\rho_e\equiv(1+p/\mu)\rho_e$ is the effective charge density serving
as the source of the electric field. This effective charge density
is state-relevant, {\it i.e.}, it is dependent on $p/\mu$. Its
significance will be discussed later.

Secondly we rewrite modified relativistic MHD in 3+1 form. Note that
the total energy-momentum tensor in $M$ can be read off from the
right hand side of Eq.~(\ref{nf}) as
\begin{equation}
T^{ab}\equiv T^{ab}_{({\mathrm{fluid}})}+\tilde
T^{ab}_{({\mathrm{em}})}+T^{ab}_{(\phi)}\ , \end{equation} where
\begin{eqnarray}
\tilde T^{ab}_{({\mathrm{em}})}&\equiv &\phi^{3/2}
T^{ab}_{({\mathrm{em}})}=\phi^{3/2}(F^{ac}F^b_{~c}-\frac14 g^{ab}F^{cd}F_{cd}) \ ,\\
T^{ab}_{(\phi)}&\equiv&-\frac{1}{k}
(g^{ab}\nabla^c\nabla_c\sqrt{\phi} -\nabla^a\nabla^b\sqrt{\phi})\
.\end{eqnarray} It is straightforward to see that \cite{gravity}
 \begin{equation}
\nabla_b T^{ab}_{({\mathrm{em}})}=F^{ac}\nabla^bF_{bc}\ .
\end{equation}
In the
light of Eqs.\ (\ref{first})-(\ref{seventh}), we obtain the 3+1 form
\begin{eqnarray}\label{em1}
\nabla_bT^{ab}_{({\mathrm{em}})}&=&n^a(-KE^2+\alpha^{-1}E_b(\mathcal{L}_t-\mathcal{L}_{\beta})E^b
-\alpha^{-1}\epsilon^{bcd}E_bD_c(\alpha B_d))\nonumber\\
&&-E^aD_bE^b+\epsilon^{abc}B_c\alpha^{-1}(\mathcal{L}_t-\mathcal{L}_\beta-\alpha
K)E_b \nonumber \\
 &&+\alpha^{-1}B_bD^a(\alpha B^b) -\alpha^{-1}B_bD^b(\alpha
B^a) \ ,
\end{eqnarray}
\begin{eqnarray}\label{em2}
T^{ab}_{({\mathrm{em}})}\nabla_b\phi^{3/2}&=&\frac{3\sqrt{\phi}}{2}\left(\frac12(B^2+E^2)(D^a\phi
+n^a\alpha^{-1}(\partial_t-\mathcal{L}_{\beta})\phi)
\right.\nonumber\\
&&-(E^aE^b+B^aB^b)D_b\phi\nonumber\\
&&\left.+E_cB_d(n^a\epsilon^{bcd}D_b\phi+\epsilon^{acd}
\alpha^{-1}(\partial_t-\mathcal{L}_{\beta})\phi) \right),
\end{eqnarray}
 here $E^2\equiv E_aE^a$ and
$B^2\equiv B_aB^a$. Recall that the relation between
three-dimensional Riemann tensor $^3R_{abc}^d$ and 4D one
$R_{abc}^d$ reads
\begin{equation}
^3R_{abc}^d=\gamma_a^e
\gamma_b^f\gamma_c^l\gamma_m^dR_{efl}^m-2K_{c[a}K_{b]}^{~d} .
\end{equation}
Using
\begin{equation}
(\nabla_a\nabla_b-\nabla_b\nabla_a)\nabla^c\sqrt{\phi}=-R_{abd}^{~~~c}\nabla^d\sqrt{\phi},
\end{equation}
a lengthy but straightforward calculation gives
\begin{eqnarray}\label{tphi}
\nabla_bT^{ab}_{(\phi)}&=&\frac1{2k\sqrt{\phi}}\left.(^3R^{ab}D_b\phi
+\alpha^{-1}(D_bK^{ab}-D^aK)(\partial_t-\mathcal{L}_{\beta})\phi\right.\nonumber\\
&&-(D^b\phi)(\alpha^{-1}(\mathcal{L}_t-\mathcal{L}_{\beta})K^a_{~b}+K^{ac}K_{bc}\nonumber\\
&&+D^aD_b\ln\alpha+(D^a\ln \alpha)D_b\ln\alpha)\nonumber\\
&&+n^a(\alpha^{-1}((\partial_t-\mathcal{L}_{\beta})\phi)(-K^{bc}K_{bc}+D^bD_b\ln\alpha+a^2\nonumber\\
&&+\alpha^{-1}(\partial_t-\mathcal{L}_{\beta})K)+(D_bK^{bc})D_c\phi-(D^bK)D_b\phi\nonumber\\
&&\left.-K_{bc}(D^b\phi)D^c\ln\alpha)\right),
\end{eqnarray}
where $^3R^{ab}$ is the three-dimensional Ricci tensor and
$a^2\equiv a^aa_a$. For a perfect fluid, the energy-momentum tensor
$T^{ab}_{({\mathrm{fluid}})}$ can also be written as
\begin{equation}
T^{ab}_{({\mathrm{fluid}})}=\rho h v^av^b+pg^{ab} \
,\end{equation}where $\rho$ is the rest-mass density as observed by
an observer co-moving with the fluid $v^a$, $p$ is the pressure and
$h$ the specific enthalpy
\begin{equation}
h=1+\epsilon+p/\rho . \end{equation} Hence one has
$\mu=(1+\epsilon)\rho$. The local conservation of the 4D Einstein
tensor $G^{ab}$ leads to
\begin{equation}\label{conservation}
\nabla_b(T^{ab}/\sqrt{\phi})=0 \ .
\end{equation}
We assume that $T^{ab}_{(\phi)}$ does not contribute to the number
of baryons. Thus we have the conservation of baryons as
\begin{equation}\label{baryon}
\nabla_a(\rho v^a)=0 \ , \end{equation}which is decomposed into 3+1
form as
\begin{equation}\label{baryontwo}
D_a(\rho \tilde
v^a)+\alpha^{-1}(\partial_t-\mathcal{L}_{\beta})(\rho W)-\rho
WK+\rho\tilde v^a D_a\ln \alpha=0 ,\end{equation} where $\tilde
v^a\equiv v^a-Wn^a$. The equation for the conservation of energy is
obtained by contracting Eq.\ (\ref{conservation}) with $n_b$ as
\begin{eqnarray}\label{energy}
H&=&\alpha^{-1}(\partial_t-\mathcal{L}_{\beta})p-W\rho(\tilde
v^aD_a h+W\alpha^{-1}(\partial_t-\mathcal{L}_{\beta})h)\nonumber\\
&&-\rho h(W\alpha^{-1}(\partial_t-\mathcal{L}_{\beta})W+\tilde
v^aD_aW\nonumber\\
&&+W\tilde v^aD_a\ln\alpha-K_{ab}\tilde v^a \tilde
v^b),\end{eqnarray}
and the Euler equation is obtained by projecting
Eq.\ (\ref{conservation}) onto $\Sigma$ as
\begin{eqnarray}\label{Euler}
\rho h\tilde v^aD_b\tilde v^a&=&\rho h(2WK^{ab}\tilde
v_b-W^2D^a\ln \alpha)-\rho hW\alpha^{-1}(\mathcal{L}_t-\mathcal{L}_{\beta})\tilde v^a\nonumber\\
&&-\rho\tilde
v^a(W\alpha^{-1}(\partial_t-\mathcal{L}_{\beta})h+\tilde v^bD_b
h)-D^ap-M^a
 ,\end{eqnarray} where
\begin{eqnarray}
H&\equiv&\phi^{3/2}(-KE^2+\alpha^{-1}E_a(\mathcal{L}_t-\mathcal{L}_{\beta})E^a-\alpha^{-1}
\epsilon^{abc}E_aD_b(\alpha B_c))\nonumber\\
&&+\frac{\sqrt{\phi}}{2}(B^2+E^2)\alpha^{-1}(\partial_t-\mathcal{L}_{\beta})\phi
+\sqrt{\phi}\epsilon^{abc}E_aB_bD_c\phi\nonumber\\
&&+\frac1{2k\sqrt{\phi}}(\alpha^{-1}(\partial_t-\mathcal{L}_{\beta})\phi(-K^{ab}K_{ab}+D^aD_a\ln\alpha+a^2\nonumber\\
&&+\alpha^{-1}(\partial_t-\mathcal{L}_{\beta})K)+(D_aK^{ab})D_b\phi-(D^aK)D_a\phi)\nonumber\\
&&-\frac {W}{2\phi}(\mu+p)\tilde v^b
D_b\phi+\frac1{2k\phi}(D_a\phi)(\alpha^{-1}(\mathcal{L}_t-\mathcal{L}_{\beta})D^a\sqrt{\phi}\nonumber\\
&&-K_{ab}D^b\sqrt{\phi})+\frac1{2\phi}\alpha^{-1}(-W^2(\mu+p)+p-\frac1k(D^aD_a\sqrt{\phi}\nonumber\\
&&+K\alpha^{-1}(\partial_t-\mathcal{L}_{\beta})\sqrt{\phi}+(D_a\sqrt{\phi})D^a\ln\alpha))(\partial_t
-\mathcal{L}_{\beta})\phi
,\end{eqnarray} and

\begin{eqnarray}
M^a&\equiv&+\sqrt{\phi}(\frac12(B^2+E^2)D^a\phi-(E^aE^b+B^aB^b)D_b\phi\nonumber\\
&&+\epsilon^{abc}E_bB_c\alpha^{-1}(\partial_t-\mathcal{L}_{\beta})\phi)-\phi^{3/2}E^aD_bE^b\nonumber\\
&&+\phi^{3/2}\alpha^{-1}(\epsilon^{abc}B_c(\mathcal{L}_t-\mathcal{L}_\beta-\alpha K)E_b+B_bD^a(\alpha B^b)
-B_bD^b(\alpha B^a))\nonumber\\
&&+\frac1{2k\sqrt{\phi}}\left(^3R^{ab}D_b\phi+\alpha^{-1}(D_bK^{ab}-D^aK)(\partial_t
-\mathcal{L}_{\beta})\phi\right)\nonumber\\
&&-(D^b\phi)\left(\alpha^{-1}(\mathcal{L}_t-\mathcal{L}_{\beta})K^a_{~b}+K^{ac}K_{bc}+D^aD_b\ln\alpha
+(D^a\ln\alpha)D_b\ln\alpha\right)\nonumber\\
&&-\frac{1}{2\phi}(\mu+p)(\tilde
v^bD_b\phi+W\alpha^{-1}(\partial_t-\mathcal{L}_{\beta})\phi)\tilde
v^a-\frac{1}{2\phi}pD^a\phi\nonumber \\
&&+\frac{1}{2k\phi}(D^a\phi)\left(D^bD_b\sqrt{\phi}-(\alpha^{-1}(\partial_t
-\mathcal{L}_{\beta}))^2\sqrt{\phi}\right.\nonumber\\
&&\left.+K\alpha^{-1}(\partial_t-\mathcal{L}_{\beta})\sqrt{\phi}+(D_b\sqrt{\phi})D^b\ln\alpha \right)\nonumber\\
&&-\frac{1}{2k\phi}\left((D_b\phi)D^aD^b\sqrt{\phi}+(\alpha^{-1}(\partial_t-\mathcal{L}_{\beta})\sqrt{\phi})
K^{ab}D_b\phi\right.\nonumber\\
&&\left.-(\alpha^{-1}(\partial_t-\mathcal{L}_{\beta})\phi)
D^a(\alpha^{-1}(\partial_t-\mathcal{L}_{\beta}) \sqrt{\phi})\right),
\end{eqnarray}
here $(\alpha^{-1}(\partial_t-\mathcal{L}_{\beta}))^2\sqrt{\phi}$
denotes
$\alpha^{-1}(\partial_t-\mathcal{L}_{\beta})(\alpha^{-1}(\partial_t-\mathcal{L}_{\beta})\sqrt{\phi})$.
Note that Eqs.\ (\ref{baryontwo}), (\ref{energy}) and (\ref{Euler})
comprise the basic formulas for the modified relativistic MHD in
three-dimensional form. In the special case $\phi\equiv1$, we have

\begin{eqnarray}
H&=&-KE^2+\alpha^{-1}E_a(\mathcal{L}_t-\mathcal{L}_{\beta})E^a-\alpha^{-1}\epsilon^{abc}
E_aD_b(\alpha B_c),\\
M^a&=&-E^aD_bE^b+\alpha^{-1}\epsilon^{abc}B_c(\mathcal{L}_t-\mathcal
{L}_\beta-\alpha K)E_b\nonumber\\
&&+\alpha^{-1}\left(B_bD^a(\alpha B^b)-B_bD^b(\alpha B^a)\right),
\end{eqnarray}
which accord with the conventional form \cite{Baumgarte}.

\section{Discussion on the effective charge}
\begin{figure}
\centering
\includegraphics[width=8cm,height=8cm]{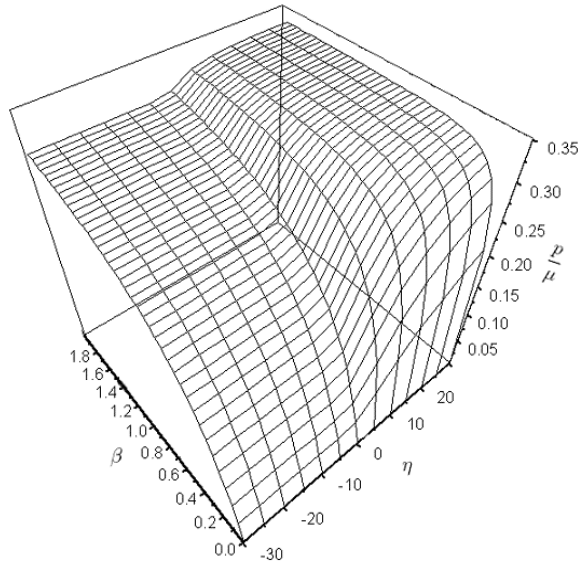}
\caption{\small  Modified term $p/\mu$ as a function of degeneracy
parameter $\eta$ and relativistic parameter $\beta$.} \label{3d}
\end{figure}

From the state-relevant Maxwell's equation~(\ref{Max}) and its 3+1
form Eq.~(\ref{M1}), we can see that $\tilde
\rho_e\equiv(1+p/\mu)\rho_e$ is the effective charge density of a
perfect fluid. Such effective charge density is relevant to the
equation of state of the fluid and its  rationality should be
carefully checked. In this section, we adopt an ideal Fermi gas as
an example to see under what condition can the modified term $p/\mu$
show visible effect. We employ in terms of the dimensionless
degeneracy and temperature parameters
\begin {equation}
\eta=\frac{\tilde{\mu}}{k_BT}, \ \ \ \beta=\frac{k_BT}{mc^2},
\end{equation}
where $\tilde{\mu}$ is the chemical potential, $m$ is the mass of
the fermion, and $k_B$ is the Boltzmann constant. The gas is
degenerate for $\eta\gg 0$ while nondegenerate for $ \eta\ll 0$. On
the other hand, the gas is extremely relativistic for $\beta\gg 1$
while nonrelativistic for $ \beta\ll 1$ \cite{fermi}. The zero of
energy for the particles is chosen so that the thermodynamic
potential reads
\begin{equation}
\Omega=-Vk_BT \int \frac{gd^3\tilde p}{h^3}
\ln\left[1+\exp\frac{\tilde{\mu}-\varepsilon}{k_BT}\right],
\end{equation}
where $\tilde p$ is the momentum, $g$ is the statistical weight, and
\begin{equation}
\varepsilon=\sqrt{(\tilde p c)^2+(mc^2)^2} -mc^2
\end{equation}
is the kinetic energy. The number density $n$, pressure $p$, and
internal energy density $E$ (per volume) of an ideal Fermi gas are
respectively
\begin{eqnarray}
n&=&K\beta^{3/2}\left[F_{1/2}(\eta,\beta)+\beta F_{3/2}(\eta,\beta)\right],\\
p&=&mc^2K\beta^{5/2}\left[\frac23 F_{3/2}(\eta,\beta)+\frac 13
\beta F_{5/2}(\eta,\beta)\right],\\
E&=&mc^2K\beta^{5/2}\left[F_{3/2}(\eta,\beta)+\beta F_{5/2}(\eta
,\beta)\right],
\end{eqnarray}
where $K=4\sqrt 2\pi g(mc/h)^3$, and the Fermi integral is
\begin{equation}
F_k(\eta, \beta)\equiv \int_0^{+\infty} \frac{z^k(1+\frac12 \beta
z)^{1/2}dz}{e^{z-\eta}+1} \ \ \ \ (k>-1).
\end{equation}
Then we have
\begin{equation}
\frac p
\mu=\frac{p}{E+nmc^2}=\frac{1}{3+{3F_{1/2}(\eta,\beta)}/({2\beta
F_{3/2}(\eta,\beta)+\beta^2F_{5/2}(\eta,\beta)})}.
\end{equation}
The relation between $p/\mu$ and $(\eta, \beta)$ is shown in
Fig.~\ref{3d}. One can see clearly that both degenerate and
relativistic conditions can lead to the value of $p/\mu$ comparable
to $1/3$ (which is the value of $p/\mu$ for radiation).

\begin{figure}
\centering
\includegraphics[width=8cm,height=6cm]{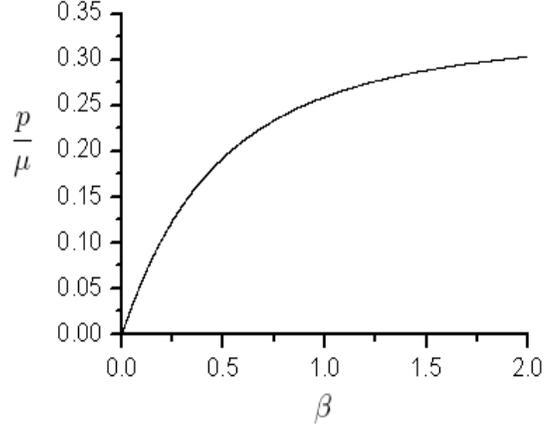}
\caption{\small  Modified term $p/\mu$ as a function of relativistic
parameter $\beta$ in nondegenerate condition $\eta=-30$.}
\label{beta}
\end{figure}

\begin{figure}
\centering
\includegraphics[width=8cm,height=6cm]{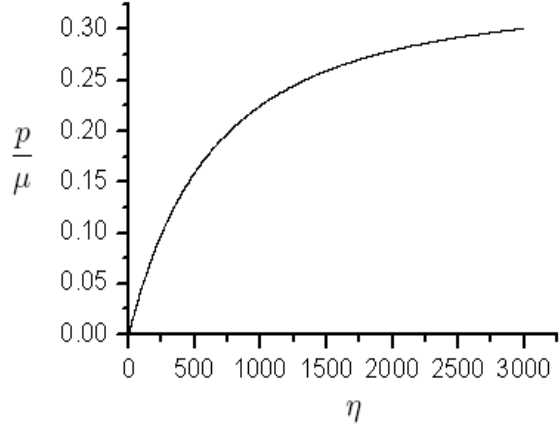}
\caption{\small  Modified term $p/\mu$ as a function of degeneracy
parameter $\eta$ in nonrelativistic condition $\beta=10^{-9}$.}
\label{eta}
\end{figure}
Now we study the two kinds of conditions respectively. For a
nondegenerate ideal Fermi gas (for example $\eta=-30$), the value of
$p/\mu$ is drawn from nonrelativistic ($\beta=0$) to relativistic
($\beta=2$) regime in Fig.~\ref{beta}. It is obvious that we need
not to go to extremely relativistic condition since $p/\mu$ is
already close to $1/3$ when $\beta=2$. In the specific calculation
for an electron gas, we set $p/\mu=0.002$ when $k_BT=1
~\mathrm{keV}$. This result indicates the possibility to test the
theory in earth laboratory. For a non-degenerate electron gas at
$T=273 K$, one may estimate the modification term as
$p/\mu \sim k_B T / m_e c^2 \sim 10^{-8}$. On the other hand, in
the experiments on the equality
of the electric charges of proton and electron, these charges in
a conductor are found to be equal within $10^{-19}$ or better
(see e.g.~\cite{Stover}). However, the proton system in a conductor
cannot be seen as a perfect fluid and hence does not satisfy our
premise. Hence the effective charges of protons in a conductor
cannot be directly obtained by our modified equations. So, those
experiments are not in severe contradiction with the KK theory.
For similar reason, the experiments reported in Ref.\cite{Gillies}
cannot provide definite opponent evidence to the KK theory either.
But this kind of experiments do cast some doubts on the classical
KK theory. Note that both the electron system and the ion system
could be regarded as perfect fluid in high-temperature plasma.
In a thermal equilibrium
state the electron and ion in a plasma have the same temperature.
Hence they would have different values of $p/\mu$. Actually the
value of $p/\mu$ for ion is much smaller than the one for electron
when $k_BT$ takes value from $\mathrm{keV}$ to $\mathrm{ MeV}$. It
turns out that the two important physical parameters for the
description of plasma--Debye length and plasma frequency \cite{PP}
have to be modified in our 5D theory as
\begin{equation}
\lambda_D=\left[\frac{\epsilon_0 k_B T}{n_e e^2
(1+p_e/\mu_e)}\right]^{\frac 12} \ ,
\end{equation}
and
\begin{equation}
\omega_{p}=\left[\frac{n_e e^2
(1+p_e/\mu_e)}{m_e\epsilon_0}\right]^{\frac 12}\ ,
\end{equation}
where $n_e$ is the number density of electron and $\epsilon_0$ the
permittivity of vacuum. Since the electromagnetic wave whose
frequency is lower than $\omega_{p}$ will be reflected while others
can transmit through the plasma, the plasma frequency can be
measured accurately~\cite{Xu}. Therefore it is possible to test the
prediction from the 5D KK theory in earth laboratory. For a
degenerate idea Fermi gas, the relation between $p/\mu$ and $\eta$
is demonstrated in Fig.~\ref{eta}. Recall that the white dwarf is
known to resist the gravity by an electronic degenerate pressure. It
is also possible to test the 5D theory by certain relevant phenomena
in outerspace.

Note that the vacuum polarization in quantum electrodynamics (QED)
also leads to an effective charge of a point-like
particle~\cite{Greiner,Peskin}. So the effective charge viewpoint
does not merely come from the KK theory. For the Fermi gas in KK
theory, the larger the density and the temperature, the larger the
effective charge factor $\tilde\rho_e/\rho_e$, which approaches to
$4/3$ as a limit. Whereas for QED, the higher the energy scale (or
shorter distance), the larger the effective charge
$e_{\mathrm{eff}}/e$, which approaches to infinity as a limit.
Therefore the state-relevant Maxwell's equation and QED give similar
results of larger effective charges. However, the state-relevant
effect in KK theory is a pure classical effect due to the extra
dimension of spacetime, whereas the QED effect is a quantum effect
irrespective of any extra dimension. So one does not expect them to
be the same. It is easy to distinguish the two effects by comparing
their characters.

In Summary, the coupling of 5D perfect fluid to KK gravity is fully
studied. The 4D effective equations of this 5D coupling system are
derived. In particular, the modified Maxwell's equation which is
relevant to the equation of state of the source is obtained. To
facilitate applications, we also derive the 3+1 form of the modified
Maxwell's equations and the relativistic MHD. It turns out that the
effective charge density in the KK theory can be written as
$\tilde\rho_e\equiv (1+p/\mu)\rho_e$. Moreover, using an ideal Fermi
gas model, we study the modification term $p/\mu$ as a function of
degeneracy parameter $\eta$ and the relativity parameter $\beta$. It
reveals that the traditional Maxwell's equation is the low density
and low temperature limit of the state-relevant Maxwell's equation.
We thus indicate the possibility to test the state-relevant effect
both in earth laboratory and in astrophysical phenomena.

\section*{Acknowledgments}
We acknowledge the valuable discussions with Lingzhen Guo, Wenan Guo, Zhi-Qiang
Guo, Bin Wu and Ren-Xin Xu. This work is supported in part by
Hui-Chun Chin and Tsung-Dao Lee Chinese Undergraduate Research
Endowment (Chun-Tsung Endowment) at Peking University, by NSFC (Nos.~10675019, 10421503, 10575003), and
by the Key Grant Project of Chinese Ministry of Education
(No.~305001).

\end{document}